\documentstyle[emulateapj,danonecolfloat]{article}
\input psfig.sty

\def\spose#1{\hbox to 0pt{#1\hss}}
\def\simlt{\mathrel{\spose{\lower 3pt\hbox{$\mathchar"218$}}
     \raise 2.0pt\hbox{$\mathchar"13C$}}}
\def\simgt{\mathrel{\spose{\lower 3pt\hbox{$\mathchar"218$}}
     \raise 2.0pt\hbox{$\mathchar"13E$}}}
\def\simpropto{\mathrel{\spose{\lower 3pt\hbox{$\mathchar"218$}}
     \raise 2.0pt\hbox{$\propto$}}}
\newcommand\lsim{\mathrel{\rlap{\lower4pt\hbox{\hskip1pt$\sim$}}
        \raise1pt\hbox{$<$}}}
\newcommand\gsim{\mathrel{\rlap{\lower4pt\hbox{\hskip1pt$\sim$}}
        \raise1pt\hbox{$>$}}}

\begin{document}
\twocolumn[



\submitted{Submitted to ApJ}

\title{The Thermal Memory of Reionization History}
\author{Lam Hui$^{1,2}$ and Zolt\'an Haiman$^3$}
\affil{$^1$ Theoretical Astrophysics, Fermi National Accelerator Laboratory, Batavia, IL 60510\\ 
$^2$ Department of Astronomy and Astrophysics, University of Chicago, IL 60637\\
$^3$ Department of Astronomy, Columbia University, New York, NY 10027\\
Electronic mail: {\tt lhui@fnal.gov}, {\tt zoltan@astro.columbia.edu}}

\begin{abstract}
The recent measurement by {\it WMAP} of a large electron scattering optical
depth $\tau_e = 0.17 \pm 0.04$ is consistent with a simple model of
reionization in which the intergalactic medium (IGM) is ionized at
redshift $z \sim 15$, and remains highly ionized thereafter.
Here, we show that existing measurements of the IGM temperature from
the Lyman-alpha (Ly$\alpha$) forest at $z \sim 2 - 4$ rule out this
``vanilla'' model. Under reasonable assumptions about the ionizing
spectrum, as long as the universe is reionized before $z = 10$ {\it
and remains highly ionized thereafter}, the IGM reaches an asymptotic
thermal state which is too cold compared to observations.  To
simultaneously satisfy the CMB and Ly$\alpha$ forest constraints, the
reionization history must be complex: reionization begins early at
$z \gsim 15$, but there must have been significant (order unity)
changes in fractions of neutral hydrogen and/or helium at $6 < z <
10$, and/or singly ionized helium at $4 < z < 10$.  We describe a
physically motivated reionization model that satisfies all current
observations.  We also explore the impact of a stochastic reionization
history and show that a late epoch of (HeII$\rightarrow$HeIII)
reionization induces a significant scatter in the IGM
temperature, but the scatter diminishes with time quickly. 
Finally, we provide an analytic formula for the thermal
asymptote, and discuss possible additional heating mechanisms that
might evade our constraints.
\end{abstract}


]

\section{Introduction}
\label{intro}

The detection by the Wilkinson Microwave Anisotropy Probe ({\it WMAP})
of a large optical depth $\tau_e$ to electron scattering has opened a
new window in studies of the ultra--high redshift ($z\sim 15$)
universe (Kogut et al. 2003, Spergel et al. 2003). Taking the value
$\tau_e = 0.17 \pm 0.04$, inferred from the polarization and
temperature anisotropies of the cosmic microwave background (CMB), at
face value, implies the reionization of the universe must begin very
early.

The optical depth to electron scattering is given by (e.g. Dodelson 2003)
\begin{eqnarray}
\tau_e = \int_0^\infty {dz \over (1+z) H(z)} \sigma_T n_e (z)
\end{eqnarray}
where $H(z)$ is the Hubble parameter at redshift $z$, $\sigma_T$ is
the Thompson cross-section, and $n_e (z)$ is the proper free electron density.
This can be rewritten as
\begin{eqnarray}
\label{taucmb}
&& \tau_e = 0.0691 \times \Omega_b h \int_0^\infty {(1+z)^2 dz \over \sqrt{\Omega_m (1+z)^3 + 
\Omega_\Lambda}} \\ \nonumber 
&& [(1 - Y_P)  X_{\rm HII} + {1\over 4} Y_P (X_{\rm HeII} + 2 X_{\rm HeIII})]
\end{eqnarray}
where $\Omega_b$, and $\Omega_m$ are the baryon and matter densities
in fraction of the critical, $h$ is the Hubble constant in units of
$100$ km/s/Mpc, and $Y_P = 0.244 \pm 0.002$ is the helium mass
fraction (Burles, Nollett \& Turner 2001).  The fractions of ionized
hydrogen $X_{\rm HII}$, singly ionized helium $X_{\rm HeII}$, and
doubly ionized helium $X_{\rm HeIII}$, are functions of $z$.

Ignoring helium, the observed $\tau_e = 0.17 \pm 0.04$ is consistent
with a universe in which $X_{\rm HII}$ changes from essentially zero
to close to unity at $z = 17 \pm 3$, and $X_{\rm HII} \sim 1$ since
(Kogut et al. 2003).  If helium is fully ionized together with
hydrogen, the reionization redshift changes slightly to $z = 15.3 \pm
2.7$.
\footnote{Throughout this paper, we adopt $\Omega_b h^2=0.024$,
$\Omega_m h^2=0.14$, and $h=0.72$ the central best--fit values
measured by WMAP (Spergel et al. 2003).}

Two features are noteworthy. First, helium reionization has a
sub-dominant effect on $\tau_e$ compared to hydrogen. Second, since the
electron scattering optical depth is controlled by the free electron
density, it is insensitive to the neutral fractions of hydrogen
($X_{\rm HI}=1-X_{\rm HII}$) and helium ($X_{\rm HeI}=1-X_{\rm HeII}-X_{\rm
HeIII}$), as long as they are small.

In contrast, the (hydrogen) Lyman-alpha (Ly$\alpha$) optical depth,
inferred from the spectra of distant quasars, is extremely sensitive
to small amounts of neutral hydrogen. The Ly$\alpha$ optical depth at
mean density is (Gunn \& Peterson 1965)
\begin{eqnarray}
\tau_\alpha = 41.8 \left[{X_{\rm HI} \over 10^{-4}}\right] \left[{1+z \over 7}\right]^3
\left[{H(z=6) \over H(z)}\right],
\end{eqnarray}
Whether $X_{\rm HI}$ is
$10^{-4}$ or $10^{-5}$, for instance, makes a big difference to
$\tau_\alpha$.  Using a model for large scale structure in the
intergalactic medium (IGM), the observed mean Ly$\alpha$ transmission
at $z \sim 6$ (or the lack thereof i.e. the Gunn-Peterson trough;
Becker et al. 2001) implies a ($1 \sigma$) lower limit on the hydrogen
neutral fraction for a fluid element at mean density: $X_{\rm HI} >
10^{-4}$.  The analog for Ly$\beta$ provides a stronger constraint,
due to the smaller absorption cross-section: $X_{\rm HI} > 5 \times
10^{-4}$ (taken from Lidz et al. 2002; see also Cen \& McDonald 2002,
Fan et al. 2002).\footnote{ Note that some authors quote volume-- or
mass--weighted neutral fractions.  We here use the neutral fraction
for a fluid element that happens to be at the cosmic mean density,
motivated by the fact that we will discuss the temperature evolution
of fluid elements with the same property in this paper.}

Taken at face value, these two separate observations are therefore
consistent with the simplest ``vanilla'' model in which the universe
is reionized in a single step: the neutral fraction experiences a
drop of order unity at at $z \gsim 15$, and it remains $\ll 1$ thereafter.
Our goal in this paper is to confront this model with
other existing observations, and to see if additional constraints can
be put on the reionization history of our universe.

Several authors have pointed out that the {\it evolution} of the
Ly$\alpha$ optical depth suggests reionization might take place not
much earlier than $z \sim 6$ (Becker et al. 2001, Djorgovski 2001, Cen
\& McDonald 2002, Gnedin 2001, Razoumov et al. 2002; but see also
Barkana 2001, Songaila \& Cowie 2002).  This is based on an
extrapolation of the mean transmission measurements from lower
redshifts. The small number of lines of sight used (a single quasar
was employed for the measurement at $z \sim 6$; but see Fan et
al. 2003 for three new $z > 6$ sources with Gunn-Peterson troughs), 
the challenge of sky subtraction and continuum extrapolation
in Gunn-Peterson trough measurements
(see discussion in Becker et al. 2001; see also Hui et al. 2001), as well as our still
maturing understanding of radiative transfer during reionization,
motivates us to look for other clues for a late period of
reionization.

The main idea is quite simple. Reionization typically heats up the IGM
to tens of thousands of degrees, and the gas subsequently cools due to
the expansion of the universe as well as due to other processes. If
the universe was reionized early, and has stayed highly ionized
thereafter, photo--ionization heating of the gas cannot overcome the
overall cooling, and the IGM might reach too low a temperature at low
redshifts compared to observations. This idea is not new
(e.g. Miralda-Escude \& Rees 1994, Hui \& Gnedin 1997, Haehnelt \& Steinmetz 1998, 
Hui 2000, Theuns et
al. 2002). Theuns et al. (2002), in particular, deduced a limit on
the reionization redshift ($z < 9$) based on temperature measurements
by Schaye et al. (2000), assuming a quasar-like ionizing spectrum, and
that HeII reionization occurs at $z \sim 3$.  
Our objective here is to seek a formulation of this
argument that is as clean and robust as possible, that reveals clearly
the underlying assumptions, and to check the consistency with the IGM
temperatures of specific models that produce the high value of
$\tau_e$ measured by {\it WMAP}.

The Ly$\alpha$ forest temperature measurements we will use are taken
from Zaldarriaga, Hui \& Tegmark (2001; ZHT01 thereafter): $T_0 = 2.1
\pm 0.9 \times 10^4$ K at $z = 2.4$, $2.3 \pm 0.7 \times 10^4$ K at $z
= 3$, and $2.2 \pm 0.4 \times 10^4$ K at $z = 3.9$.  Note that 2
$\sigma$ errorbars are quoted, and the temperature $T_0$ was derived
for fluid elements at the mean density, consistent with our modeling of the IGM
temperature in the rest of this paper.  There have been several other
measurements in the past (Ricotti, Gnedin \& Shull 2000, Schaye et
al. 2000 [ST00], Bryan \& Machacek 2000, McDonald et al. [MM01], 
Meiksin, Bryan \& Machacek 2001). The
ones that are easiest to compare, because they are based on very
similar datasets, are ST00, MM01 and ZHT01.  The former two are based
on line width measurements, while the last one makes use of the small
scale transmission power spectrum. A virtue of the last method is that
the temperature constraints come from marginalizing over a wide array
of parameters, including the slope and amplitude of the primordial
power spectrum (e.g. Hui \& Rutledge 1999), and the equation of state index. It is reassuring that
MM01 and ZHT01, using very different methods and employing different
simulations (MM01 using hydrodynamic simulations, and ZHT01 using
N-body simulations with a marginalized smoothing to mimic Jeans
smoothing), agree well with each other.  The results of ST00 are
somewhat discrepant from these two works -- the reader is referred to
ZHT01 for further discussions.

An important issue in determining the temperature of the IGM is the
second ionization of helium (HeII$\rightarrow$HeIII). As we will see,
this can be an important source of heating at low redshifts $z\sim
3-4$.  There are several lines of evidence that suggest HeII might be
reionized at $z \sim 3$, including observations of HeII patches that
do not seem to correlate with HI absorption (Reimers et al. 1997, Anderson et
al. 1999, Heap et al. 2000, Kriss et al. 2001, Jakobsen et al. 2003), increase in IGM temperature (Schaye et
al. 2000; Theuns et al. 2002), the evolution of the hardness of the
ionizing background spectrum (Songaila 1998), and evolution of the
mean transmission (Bernardi et al. 2003).  On the other hand, it is
unclear if the fluctuations in HeII absorption observed in a few lines
of sight might not be due to a naturally fluctuating IGM
(Miralda-Escude, Haehnelt \& Rees 2000); the IGM temperature
measurements by MM00 and ZHT01 are consistent with no feature at $z
\sim 3$; power spectrum evolution also seems to argue against HeII
reionization at $z \sim 3$ (McDonald \& Seljak, private
communication). In this paper, when we consider constraints from the
temperature measurement, primarily at $z = 3.9$, we therefore leave
two options open: HeII can be ionized or not ionized by $z = 3.9$.
With this explained, we can now state our vanilla model in more
concrete terms.  It has two variants: 1. both hydrogen and helium are
fully reionized at $z \gsim 15$, and they remain highly ionized
thereafter; 2. the same as 1. except that helium is only singly
ionized, and remains so until at least past $z = 3.9$.

The paper is organized as follows. In \S \ref{asymptote}, we explain
the idea of a thermal asymptote for the IGM, and use it to derive a
constraint on the hardness of the ionizing spectrum if the universe
were to reionize before $z = 10$, and remains highly ionized
thereafter. In \S \ref{Jnu}, we discuss limits on the hardness of the
ionizing spectra, and show that these spectra fall short of making the
thermal asymptote sufficiently hot to match observations.  We then
work out illustrative examples in \S \ref{reheat} of how order unity
changes in the neutral fractions at $z \lsim 10$ can reproduce the
temperature measurements, while being consistent with the {\it WMAP}
data.  We go on to offer a physically motivated model in \S
\ref{model}, and discuss the implications of the fact that different
fluid elements are reionized at different times (stochastic
reionization). We conclude in \S \ref{discuss}.

\section{Thermal Asymptotics}
\label{asymptote}

Several thermal processes are at work in a photoionized IGM. 
They are described in detail in Hui \& Gnedin (1997). Here is a brief qualitative summary:
\begin{itemize}
\item Adiabatic heating/cooling. Gas elements can heat up or cool simply due to
adiabatic contraction or expansion. The overall expansion of the universe
drives a temperature fall-off as $(1+z)^2$ as $z$ decreases.
\item Photoionization heating. Photons inject energy 
into the gas by ionizing hydrogen or helium. 
\item Recombination cooling. Protons and electrons (or ionized helium and
electrons) can cool by recombining and radiating energy away. 
\item Compton cooling. At sufficiently high redshifts, $z \gsim 10$, Compton
scattering of free electrons with the lower temperature CMB can be an important
source of cooling. 
\end{itemize}

\begin{figure}[tb]
\centerline{\epsfxsize=9cm\epsffile{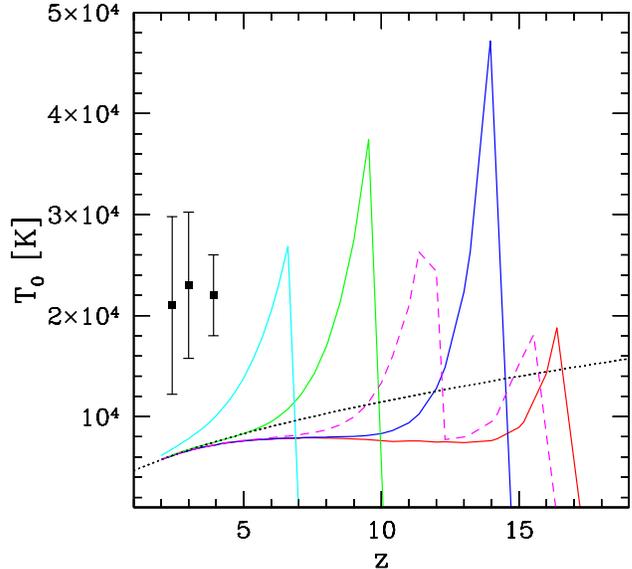}}
\caption{The figure shows the thermal asymptote (black dotted line;
eq. [\ref{analytic}]), and illustrates the fact that a wide range of
ionization histories result in the same IGM temperature by redshift
$z=4$, unless reionization occurs late.  
Each colored solid line describes the evolution of the
temperature ($T_0$) for a fluid element at mean density, according to
a different reionization history (and a different initial
reheat-temperature).  The (magenta) dashed line illustrates the
thermal evolution for a complex reionization history -- such
complexities do not stop the temperature from reaching the late time
asymptote, as long as they take place early, before $z \sim 10$.  The
points with (2 $\sigma$) error-bars on the left are measurements of
$T_0$ from the Ly$\alpha$ forest (Zaldarriaga et al. 2001).  }
\label{asymp}
\end{figure}

Typically, photoionization heating provides the dominant source of
heating for the tenuous IGM. This occurs in primarily two forms. One
is during what we call reionization 'events' -- these are (possibly
extended) periods in which a given gas element experiences order unity
changes in the fractions of HI/HeI/HeII. The other is photo-heating
for an already highly ionized plasma -- this occurs through
photoionization of small amounts of HI/HeI/HeII, whose abundances are
determined by photoionization equilibrium.  The former provides a big
boost to the temperature, while the latter determines the asymptotic
thermal state of the IGM. Typical temperature evolutions are
illustrated in several toy models in Fig. \ref{asymp}.

Fig \ref{asymp} shows the evolution in temperature $T_0$ for a fluid
element at mean density, with a variety of reionization histories.
These histories have different reionization redshifts and initial
temperature jumps, selected here only for illustration. The code for
computing the thermal and chemical evolution is described in Hui \&
Gnedin (1997).  For each thermal curve, one can see a rise to high
temperatures after a reionization event.  Thereafter, the varieties of
reionization histories and initial reheat temperatures all result in a
rather similar late time thermal asymptote (black dotted line).  In
fact, if reionization occurs before $z \sim 10$ (and the IGM stays
highly ionized since, or to use our earlier terminology: no
reionization events take place thereafter), the temperature is always
very close to the asymptote for $z < 4$, which is where temperature
measurements exist (points with error-bars).  This is true even if
reionization is not a single step process, as illustrated by the
dashed (magenta) line, as long as complexities in the reionization
process occur before $z \sim 10$. The term 'complexities' here has a
rigorous meaning: it refers to two or more episodes during which the
neutral fractions of hydrogen and/or helium undergo changes of order
unity. In all thermal curves shown in Fig. \ref{asymp}, with one
exception, such changes take place before about $z \sim 10$. The
exception, i.e. the leftmost curve where reionization takes place at
$z \sim 7$, is also the only one where the asymptote is {\it not}
reached by $z = 4$ -- this is because there is not sufficient time for
the IGM to cool after a recent episode of reionization.

Fig. \ref{asymp} therefore illustrates two very important general facts:

\begin{enumerate}
\item {\it The thermal state at $z \le 4$ does not remember the part of reionization
history prior to $z \sim 10$.}

\item {\it The IGM does, however, retain short-term memory of reionization events in its recent past.}
\end{enumerate}

What determines the late time thermal asymptote? It is the combination of photoionization heating,
recombination cooling, and adiabatic cooling. An approximate analytic
expression can be derived for the thermal evolution of the IGM (Hui \& Gnedin 1997).
We find that the following simple formula provides a more accurate ($\sim 5 \%$) fit to the 
numerically computed thermal asymptote (for $T_0$ at $z = 2 - 4$), for a variety of spectral shapes we
have tested:
\begin{eqnarray}
T_0 = [B (1 + z)^{0.9}]^{1\over 1.7}
\label{analytic}
\end{eqnarray}
where
\begin{eqnarray}
\label{BT}
&& B \equiv  
18.8 \, {\rm K}^{0.7} \, \left[\sqrt{0.14 \over \Omega_m h^2} {\Omega_b h^2 \over 0.024}\right]\\
\nonumber
&& \quad \times \left(T_{\rm HI} + {X_{\rm HeII} \over 16} T_{\rm HeI} + {5.6 X_{\rm HeIII} \over 16}
T_{\rm HeII} \right) \\ \nonumber
&& T_{i} \equiv k_B^{-1}
{\int_{\nu_{i}}^\infty J_\nu \sigma_{i} ({\rm h}\nu - {\rm h}\nu_{i}) d\nu/({\rm h} \nu) \over
\int_{\nu_{i}}^\infty J_\nu \sigma_{i} d\nu/({\rm h} \nu)} 
\end{eqnarray}

Here, $k_B$ is the Boltzmann constant, the symbol $J_\nu$ denotes the ionizing intensity
as a function of frequency (it has unit of energy per frequency per time per area
per ster-radian), $\sigma_i$ denotes the ionization cross-section for the respective
species ($i = {\rm HI, \, HeI}$ or ${\rm HeII}$), $\nu_i$ is the ionizing threshold frequency,
and ${\rm h}$ is the Planck constant (in distinction from the Hubble constant $h$). 
It is useful to remember ${\rm h}\nu_{\rm HI}/k_B = 
1.57807 \times 10^5$ K, $\nu_{\rm HeI} = 1.808 \, \nu_{\rm HI}$, and
$\nu_{\rm HeII} = 4 \, \nu_{\rm HI}$. 

The asymptote given in eq. (\ref{analytic}) assumes that at late
times, hydrogen is highly ionized. That is why the term due to
photo-heating of HI, $T_{\rm HI}$, has the form shown in
eq. (\ref{BT}).  The denominator of $T_{\rm HI}$, when multiplied by
$4 \pi$, is the photoionization rate of neutral hydrogen -- its
inverse is hence proportional to the (small) hydrogen neutral fraction
under ionization equilibrium.  Its numerator gives the photo-heating
rate per neutral atom.  The combination of factors in $T_{\rm HI}$
therefore gives a temperature scale whose amplitude is proportional to
the net amount of HI photo-heating.  Similarly, the terms $X_{\rm
HeII} T_{\rm HeI} / 16$, and $5.6 X_{\rm HeIII} T_{\rm He II} / 16$ in
eq. (\ref{BT}) quantify the importance of photo-heating of HeI and
HeII respectively.  The factors of $X_{\rm HeII}$ and $X_{\rm HeIII}$
arise due to photoionization equilibrium. If, asymptotically, helium
is doubly (singly) ionized, then $X_{\rm HeII}$ ($X_{\rm HeIII}$) can
be set to zero.

It is important to emphasize that the thermal asymptote given in eq. (\ref{analytic})
is determined completely by the shape (or 'hardness') of the ionizing spectrum $J_\nu$,
but not its amplitude. For a power-law $J_\nu \propto \nu^{-\alpha}$, the
thermal asymptote can be recast simply as:
\begin{eqnarray}
\label{T0simple}
T_0 = 2.49 \times 10^4 {\,\rm K \,} \times (2+\alpha)^{-{1\over 1.7}} \left({1+z \over 4.9}\right)^{0.53}
\end{eqnarray}
assuming, asymptotically, helium is doubly ionized. 

Given the above, we can ask the following question: how hard does the
ionizing spectrum have to be for the thermal asymptote to match the
observed temperatures at low redshifts?  It suffices to use the
measurement at the highest redshift, $z = 3.9$: $T_0 = 2.2 \pm 0.4
\times 10^4$ K (2 $\sigma$ error-bar; ZHT01).  More concretely, what
constraint can be placed on $\alpha$ if the thermal asymptote were to
reach $T_0 > 1.6 \times 10^4$ K (3 $\sigma$ lower limit), by $z =
3.9$?  It is straightforward to show that this requires $\alpha <
0.12$, from eq. (\ref{T0simple}).

One can also derive a similar limit if it is assumed helium is only
singly ionized by $z = 3.9$ (i.e. $X_{\rm HeI}, X_{\rm HeIII} \ll 1$,
$X_{\rm HeII} \sim 1$).  Assuming again a power-law $J_\nu \propto
\nu^{-\alpha}$, but with a cut-off for $\nu > \nu_{\rm HeII}$, the
requirement is $\alpha < -2.2$. In other words, the spectrum needs to
be even harder with no HeII photo-heating, and a spectrum as hard as
$\alpha < -2.2$ is unrealistic in comparison with stellar and quasar
spectra.

To summarize: {\it if reionization takes place at $z > 10$, and the
fractions of HI/HeI/HeII experience no significant (order unity)
change after $z = 10$, a sufficiently hard ionizing spectrum is
necessary to keep the IGM temperature high enough to match
observations at $z = 3.9$.} Parameterizing the ionizing background by
a power-law $J_\nu \propto \nu^{-\alpha}$, this requires $\alpha <
0.12$ if HeII is reionized by $z = 3.9$, or $\alpha < -2.2$ if HeII is
not reionized by then (the latter assumes $J_\nu$ is cut off for $\nu
> \nu_{\rm HeII}$).  One should keep in mind that the only relevant
slope of the spectrum is the slope just above each ionization
threshold (unless the spectrum is very hard), because the ionization
cross section $\sigma_i \sim \nu^{-3}$.  In other words, $\alpha$ can
deviate greatly from the values given above as long as the deviation
takes place at frequencies far away from the ionization thresholds.  Note
also that the spectrum described here refers to the asymptotic
spectrum at $z = 3.9$. If the spectrum changes significantly after $z
\sim 10$, one can view the above limits as applicable to the hardest
spectrum between $z = 3.9 - 10$.

The power index we find, $\alpha < 0.12$, or $\alpha < -2.2$,
represents a very hard spectrum. We next turn to the question: how hard can
a realistic ionizing spectrum be?

\section{The Ionizing Spectrum}
\label{Jnu}

Two kinds of ionizing spectrum are generally discussed in the literature.
One is quasar-like and the other is star-like. 

Zheng et al. (1998) finds a quasar spectral shape of $\sim \nu^{-1.8}$
for the relevant ionizing frequencies in high resolution {\it HST}
spectra.  In this paper, we will follow Haardt \& Madau (1996) and
consider, conservatively, a quasar spectrum that goes as
$\nu^{-1.5}$. One should keep in mind that, at the relevant redshifts
$\gsim 4$, the known populations of quasars probably cannot contribute
significantly to the (hydrogen) ionizing background.  Here, we take
the conservative view that there might be a population of dim quasars
that still contribute significantly to a hard spectrum (Haiman \& Loeb
1997).

Stellar spectra are generally softer than a typical quasar
spectrum. An exception is the spectrum of metal-free stars (Tumlinson
\& Shull 2000, Bromm, Kudritzki \& Loeb 2001, Schaerer 2002,
Venkatesan, Tumlinson \& Shull 2003).  In
their theoretical models, Bromm et al. (2001) find a spectrum that has
the following form: $J_\nu \sim \nu$ for $\nu$ just above $\nu_{\rm
HI}$, $J_\nu \sim \nu^0$ at $\nu \gsim \nu_{\rm HeI}$, and $J_\nu \sim
\nu^{-4.5}$ for $\nu \gsim \nu_{\rm HeII}$. Such a spectrum is harder
than the quasar spectrum at frequencies below the HeII threshold. Note
that metal free stars probably cause reionization early on, but it is
unlikely they contribute significantly to the asymptotic ionizing
spectrum at low redshifts (Haiman \& Holder 2003). We consider a
metal free stellar spectrum here for the sake of being conservative
i.e. assume a spectrum that is as hard as possible.

The actual ionizing spectrum seen by a fluid element is different from
the above, due to processing by the IGM.  Haardt \& Madau (1996) have
done a careful calculation of such effects.  Absorption generally
hardens the spectrum\footnote{Note the apparent paradox: absorption,
by taking away ionizing photons, {\it increases} the photo-heating
rate.  This is a result of photo-ionization equilibrium, which makes
the amplitude of the ionizing background irrelevant
(eq. [\ref{BT}]). Rather it is the spectral shape that is important.}
However, diffuse recombination radiation from the absorbing medium
tends to compensate for this hardening. Haardt \& Madau (1996) found
that the spectrum above $\nu_{\rm HeII}$ hardens by $1.5$ 
(to be precise, $\alpha \rightarrow \alpha - 1.5$) (see also Zuo \& Phinney 1993).  
The spectrum just
above $\nu_{\rm HI}$ and $\nu_{\rm HeI}$ maintains roughly the same
slope as the source.

We are therefore led to consider the following IGM modified spectra.
Let us use the symbols $\alpha_{\rm HI}$, $\alpha_{\rm HeI}$ and
$\alpha_{\rm HeII}$ to denote the spectral slopes (more precisely, its
negative), above the three relevant ionizing thresholds.  A
quasar-like processed spectrum has $\alpha_{\rm HI} = \alpha_{\rm HeI}
= 1.5$, and $\alpha_{\rm HeII} = 0.0$.  A metal-free-stellar spectrum
gives rise to $\alpha_{\rm HI} = -1$, $\alpha_{\rm HeI} = 0$, and
$\alpha_{\rm HeII} = 3$.  At the moment(s) of reionization, the
relevant spectrum can, however, be even harder (Abel \& Haehnelt
1999). In principle, the spectrum can be hardened relative to the
source by as much as a power-law index of $3$, the $3$ coming from the
scaling $\sigma_i \propto \nu^{-3}$ (see arguments in Abel \& Haehnelt
(1999), and also Zuo \& Phinney 1993). We therefore allow the
quasar-spectrum to have $\alpha_{\rm HI} = \alpha_{\rm HeI} =
\alpha_{\rm HeII} = -1.5$, and the stellar-spectrum to have $\alpha_{\rm
HI} = -4, \alpha_{\rm HeI} = -3, \alpha_{\rm HeII} = 1.5$, at the
initial moment(s) of reionization.  This is not relevant for the
thermal asymptote, but is relevant for the magnitude of temperature
boosts during reionization events.

Comparing the above spectral slopes against the limits obtained in the
last section suggests neither quasars, nor metal--free stars can match
the observed temperature at $z = 3.9$, if reionization occurs
at $z > 10$, and no reionization event takes place afterwards. However, those limits were
based on a strict power-law spectrum. The thermal asymptotes
(eq. [\ref{analytic}]) for our more realistic spectra for quasars and
metal--free stars are shown in Fig. \ref{quasar} and Fig. \ref{star}
(black dotted lines), respectively. Indeed, in both cases, the thermal
asymptotes fall short of the observed (3 $\sigma$) lower limit of $T_0
> 1.6 \times 10^4$ K at $z = 3.9$, regardless of whether or not helium
is doubly ionized.

Hence, if the universe is reionized before $z \sim 10$, and remains
highly ionized thereafter, neither reasonably hard spectra can
reproduce the IGM temperatures inferred from the Ly$\alpha$ forest.

\section{Illustrative Examples}
\label{reheat}

The conclusion from the last section, together with the large electron
scattering optical depth measured by {\it WMAP}, therefore implies
that one or more of the fractions $X_{\rm HI}, X_{\rm HeI}, X_{\rm
HeII}$ must change by order unity at $z < 10$, to give the IGM
temperature boosts above the thermal asymptote.  {\it We therefore
rule out the vanilla reionization models laid out in \S \ref{intro}.}
One should keep in mind, however, this is predicated upon (reasonable)
assumptions about the hardness of the ionizing spectrum.  In this
section, we work out some illustrative examples of what it takes
to match the IGM temperature constraints, using the spectra from
the last section. 

\begin{figure}[tb]
\centerline{\epsfxsize=9cm\epsffile{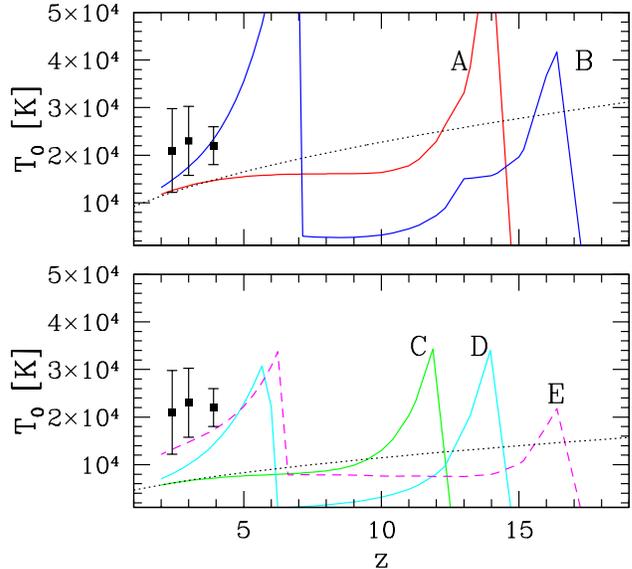}}
\caption{Thermal evolution for a quasar-like ionizing spectrum.  {\it
Upper panel: helium is doubly ionized together with hydrogen.}  The
(black) dotted line shows the thermal asymptote
(eq. [\ref{analytic}]).  The (colored) solid lines show the thermal
evolution for two different reionization histories. Model A has a
single-episode reionization at $z = 14$, and stays highly ionized
thereafter.
Model B also has early reionization, but
experiences a drop in ionizing flux thereafter, and starts recombining until $z \sim 7$
at which time it undergoes a second period of hydrogen and helium (I and II) reionization. 
Model B, but not A, reaches a high enough $T_0$ to match observations (points with error-bars; ZHT01). 
{\it Lower panel: helium is singly ionized together with hydrogen; the
double ionization of helium is never reached (except in model E}).  The
(black) dotted line shows the asymptote in such a case.  Models C, D
(colored solid lines) are analogs of A and B above -- the only
difference is that here, helium remains only singly ionized.  In model
E (magenta dashed line), hydrogen and helium is (singly) ionized at $z
\sim 17$, and then HeII is ionized at $z \sim 6.5$.  }
\label{quasar}
\end{figure}

Fig. \ref{quasar} shows the thermal evolution for a fluid element at
mean density subject to a quasar-like ionizing spectrum, and
experiencing a variety of ionization histories, for both the case of
having helium doubly ionized (upper panel), and the case of only
having singly ionized helium (lower panel; except for model E, see below).  Models A and C both
describe early reionization before $z = 10$, and no significant (order
unity) changes in $X_{\rm HI}$, $X_{\rm HeI}$, $X_{\rm HeII}$
thereafter.  As explained before, both converge to their respective
thermal asymptotes by $z \sim 4$, which fall short of the observed
temperatures, especially for model C, which has no HeII reionization,
and therefore lower temperatures.  Model B demonstrates how early
reionization can be made consistent with the forest temperature
measurements. It has a late second episode of (both hydrogen and
helium I and II) reionization at $z \sim 7$, which boosts the temperature
significantly above the thermal asymptote. Model E is similar, except
that in the second episode, only HeII is reionized (HI and HeI were
already highly ionized before then).  It shows that {\it HeII
reionization alone can provide a significant boost to the IGM
temperature.}

Model D is intriguing, as it shows a case where a second episode of
hydrogen (and HeI, but not HeII) reionization occurs at the smallest
redshift ($z = 6$) allowed by SDSS observations (Fan et al. 2002). Even
with such a late reionization epoch, the temperature can barely match
the observations (the temperature at $z = 3.9$ is $1.5 \times
10^{-4}$, which is just below the 3 $\sigma$ limit).  This is due to
the lack of HeII reionization in this model.  {\it Therefore, with a
quasar spectrum like the one assumed here, some amount of HeII
reionization is necessary prior to $z \sim 4$, to heat the IGM
sufficiently.} If the evidence that suggests that HeII reionization is occurring
at $z \sim 3$ holds up (see \S \ref{asymptote}), then this implies
either that a spectrum harder than assumed here for quasars exists
prior to $z \sim 4$, or else HeII reionization must last for a an
extended period, from $z \sim 3$ back to at least $z > 4$.

\begin{figure}[tb]
\centerline{\epsfxsize=9cm\epsffile{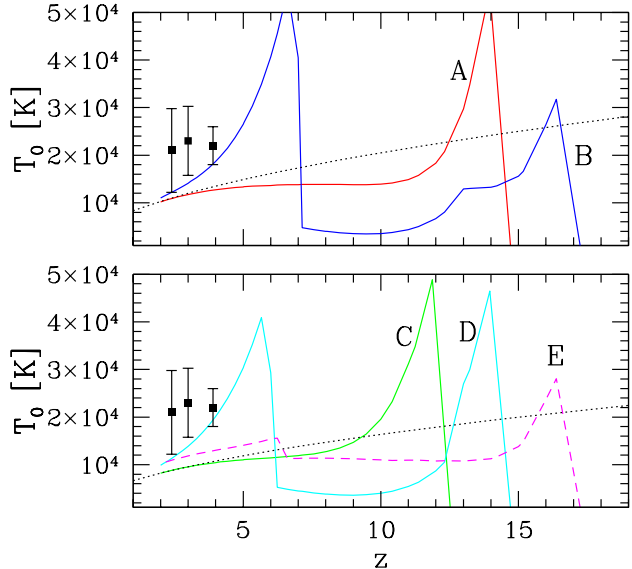}}
\caption{Thermal evolution using a spectrum motivated by metal-free stars. All
reionization histories are exact analogs of those in Fig. \ref{quasar}. All symbols
inherit the same meanings. 
}
\label{star}
\end{figure}

Fig. \ref{star} shows a similar exercise for a metal free stellar
spectrum (as defined in \S \ref{Jnu}).  We emphasize again, however,
that it is unlikely that metal free stellar spectra remain a dominant
contribution to the ionizing background down to low redshifts (Haiman
\& Holder 2003).  This figure should only be viewed as an illustration
of possibilities.  The curves are exact analogs of those in
Fig. \ref{quasar}.  Two features are different from the previous
figure.  Model E, where a late period of HeII reionization occurs
around $z \sim 6$, can no longer heat up the IGM sufficiently to be
consistent with observations.  This is because of the softness of a
stellar spectrum above the HeII threshold.  On the other hand, model
D, which has a late hydrogen (and HeI, but no HeII) reionization at $z
\sim 6$, has no problem matching the observed temperatures, unlike its
quasar analog shown in Fig. \ref{quasar}.  This is because the stellar
spectrum we adopted is, in fact, harder than the quasar spectrum for
frequencies just above $\nu_{\rm HI}$ and $\nu_{\rm HeI}$.

\begin{figure}[tb]
\centerline{\epsfxsize=9cm\epsffile{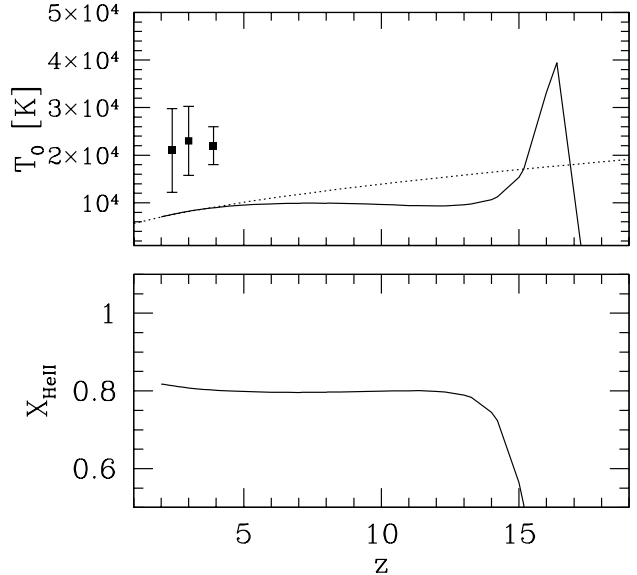}}
\caption{The thermal evolution (solid line, upper panel)
using the same quasar-like spectrum as is assumed in Fig. \ref{quasar},
except that the amplitude of the ionizing radiation just above the HeII threshold
is tuned to give rise to partial HeII ionization (lower panel). 
The dotted line shows the thermal asymptote given in eq. (\ref{analytic}).
}
\label{someHeII}
\end{figure}

Finally, before we move on to the next section, it is interesting to 
explore a situation that is intermediate between the two extremes
we have considered above i.e. HeII is partially ionized early on,
and stays partially ionized thereafter. Because the recombination rate
of HeIII to HeII is high, one might wonder if
the continuous photoionization and recombination could give rise
to a sufficiently high temperature.\footnote{We thank Andrea Ferrara for
posing the question.} We show in Fig. \ref{someHeII}
the thermal evolution (upper panel) for a fluid element
which experiences reionization at $z > 10$, with the HeII fraction (lower panel)
maintaining at the $80 \%$ level thereafter. The thermal evolution
assumes a quasar spectrum identical to that used in Fig. \ref{quasar}
(in particular, $\alpha_{\rm HeII} = 0$ for the processed, post-reionization spectrum;
see \S \ref{Jnu}), except that a break in the ionizing flux just above the HeII threshold
is tuned to give rise to $X_{\rm HeII} \sim 0.8$. 
Notice how the analytic asymptote (dotted line; from eq. [\ref{analytic}], with $X_{\rm HeII} = 0.8$, and
$X_{\rm HeIII} = 0.2$) remains an excellent approximation to the
late time thermal evolution. One can see that the late time temperature is
intermediate between the cases of full HeII reionization and of no HeII reionization, 
depicted in Fig. \ref{quasar}. In other words, partial 
HeII reionization is nicely bracketed by the cases we have considered already.
A caveat one should keep in mind, however, is that when $X_{\rm HeII}$ is close to unity,
the processed ionizing spectrum is expected to be still harder above the HeII threshold
compared to what we have already assumed.
Trying $\alpha_{\rm HeII} = -1.5$ (which is the hardest possible; see \S \ref{Jnu})
in place of $\alpha_{\rm HeII} = 0$ in Fig. \ref{someHeII}, 
results in a temperature at $z \sim 4$ of $1.3 \times 10^4$ K, which still falls
short of the observed temperature. For smaller $X_{\rm HeII}$, one expects the 
spectrum to be softer.

\section{Stochastic Reionization History -- Results from a Physically Motivated Model}
\label{model}

In the above sections we have used toy models of reionization to
illuminate the key issues that determine the temperature evolution of
the IGM. It is interesting to consider this evolution in a physically
motivated model of the reionization history that appears to fit all
the relevant observations (including the electron scattering optical
depth $\tau_e=0.17$ measured by {\it WMAP}).  Based on assumptions about the
nature and efficiency of the ionizing sources, the reionization
history can be predicted from ``first principles'' in numerical
simulations (e.g .Gnedin \& Ostriker 1997; Nakamoto, Umemura, \& Susa
2001; Gnedin 2001; Razoumov et al. 2002), and in semi--analytical
models (e.g. Shapiro, Giroux \& Babul 1994; Tegmark et al. 1994;
Haiman \& Loeb 1997, 1998; Valageas \& Silk 1999; Wyithe \& Loeb 2003;
Cen 2003).  Here we consider a semi--analytical model adopted from
Haiman \& Holder (2003), but modified to include the ionization of
HeII$\rightarrow$HeIII.  For technical details, the reader is referred
to that paper.  Inclusion of HeIII is conceptually straightforward.             

In this model, we follow the volume filling fractions $x_{\rm HII}$
and $x_{\rm HeIII}$ of HII and HeIII, assuming that discrete ionized
Str\"omgren spheres are being driven into the IGM by ionizing sources
located in dark matter halos.  As can be seen from 
eq. (\ref{BT}), photoionization of HeI (to HeII) plays a
sub-dominant role in heating the IGM. For simplicity, we assume that HeI
is reionized at the same time as HI, so there is no need to separately
keep track of the filling fraction of HeI. Note that before the
Str\"omgren spheres percolate, the radiation background is extremely
inhomogeneous: fluid elements inside ionized regions see the flux of a
single (or a cluster of a few) sources, whereas fluid elements in the
still neutral regions see zero flux.\footnote{These statements would
no longer hold if the early ionizing sources had a hard spectrum
extending to $\gsim$1 keV energies; an interesting possibility
(e.g. Oh 2001, Venkatesan, Giroux \& Shull 2001) that we do not consider in this paper.} In this
picture, each fluid element is engulfed by an ionization front at a
different time. In effect, each fluid element therefore has a
different reionization history.  Rather than considering a single
temperature for a fluid element at the mean density, it is more
appropriate to consider an {\it ensemble} of fluid elements at the
mean density, each with a different reionization history and
temperature evolution.  Note that this stochasticity is {\it in
addition} to the IGM having a distribution of temperatures due to
density variations (see \S \ref{discuss} for a discussion of the latter).

\begin{figure}[tb]
\centerline{\epsfxsize=9cm\epsffile{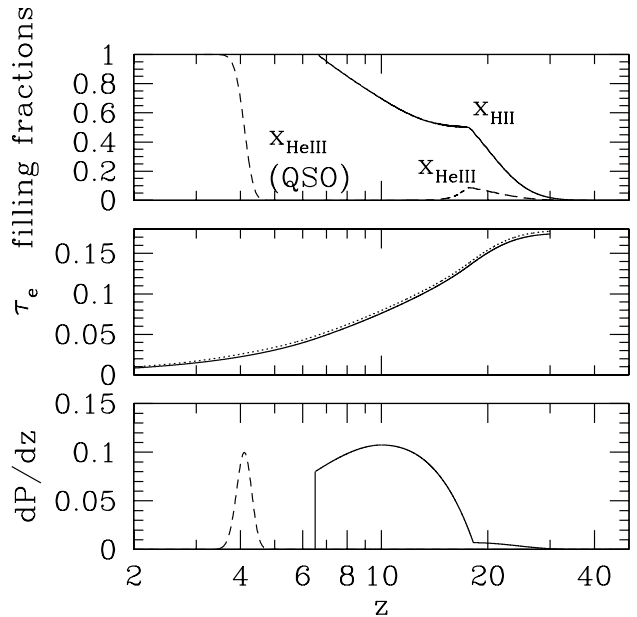}}
\caption{{\it Top panel.} The evolution of the volume filling
fractions $x_{\rm HII}$ and $x_{\rm HeIII}$ of ionized hydrogen and
helium in a physically motivated reionization model. The solid curve
corresponds to HII, and the dashed curve to HeIII regions. The extra
dashed curve to the left (QSO) corresponds to the evolution of volume
filling fraction of HeIII regions due to a late period of HeII
reionization, driven by the population of known quasars at $z>3$.
{\it Middle panel.}
The electron scattering optical depth, integrated
from $0$ to $z$.  Two barely distinguishable curves are shown here:
excluding/including free electrons from the quasar-induced HeII
ionization.  {\it Bottom panel.} The solid curve shows the probability
distribution of hydrogen reionization redshifts. The dashed curve
shows the distribution of (late) HeII reionization redshifts. The
amplitude of the
latter distribution has been divided by a factor 20, for
clarity of presentation.
}
\label{fig:bubbles}
\end{figure}

The evolutions of $x_{\rm HII}$ and $x_{\rm HeIII}$ in our model are
shown by the solid and dashed curves in the top panel of
Figure~\ref{fig:bubbles} (ignore the QSO curve to the left for the
moment).  Reionization has an interesting history that reflects
contributions from three distinct types of ionizing sources, and two
different feedback effects (all of which have physical motivations as
described in detail in Haiman \& Holder 2003).  In short, ionizing
sources (assumed to be massive metal--free stars) first appear inside
gas that cools via ${\rm H_2}$ lines, and collects in the earliest
non--linear halos with virial temperature of $100\,{\rm K} \lsim T
\lsim 10^4$K.  These sources ionize $\sim50\%$ of the volume in
hydrogen, and they have sufficiently hard spectra (see discussion
above) that they reionize $\sim10\%$ of the helium. However, at this
stage (redshift $z\sim 17$) the entire population of these first
generation sources effectively shuts off due to global ${\rm
H_2}$--photodissociation by the cosmic soft UV background they had
built up.  Soon more massive halos, with virial temperatures of
$10^4\,{\rm K} \lsim T \lsim 2\times 10^5$K form, which do not rely on
${\rm H_2}$ to cool their gas (they cool via neutral H excitations),
and new ionizing sources turn on in these halos.  These are assumed to
be ``normal'' stars, since the gas had already been enriched by heavy
elements from the first generation.  These sources continue ionizing
hydrogen, but since they produce little flux above the HeII edge,
helium starts recombining. This second generation population is also
self--limiting: gas infall to these relative shallow potential wells
is prohibited inside regions that had already been ionized and
photo--heated to $10^4$K.  As a result, hydrogen reionization starts
slowing down around $z\sim 10$ (see the solid curve in the bottom
panel of Fig.~\ref{fig:bubbles}). However, at this stage, still larger
halos with virial temperatures of $T \gsim 2\times 10^5$K start
forming. These relatively massive, third generation halos are
impervious to photoionization feedback, and complete the reionization
of hydrogen.

As far as the late time thermal state is concerned, the relevant
ionizing spectrum is that of the sources that turn on after $z \sim
17$. These are Population II stars -- a reasonable spectrum is
$\alpha_{\rm HI} = 1$, and $\alpha_{\rm HeI} = 4$, and heavily
truncated beyond $\nu_{\rm HeII}$ (Leitherer et al. 1999). As we have
illustrated with examples in the last section, a spectrum as soft as
this, even with some amount of late (HI, HeI but not HeII)
reionization after $z \sim 10$, has difficulty heating up the IGM to
high enough temperatures (even after accounting for processing of
spectrum by the IGM, see \S \ref{Jnu}).

We are therefore led to consider the effect of quasars.  Sokasian,
Abel \& Hernquist (2002) showed, based on a 3D radiative transfer
simulation, that HeII reionization should occur
around $z \sim 4$, driven by the population of known
quasars. 
\footnote{Wyithe \& Loeb (2002) considered HeII reionization
by the known quasar population in a semi-analytical model and
obtained a similar result: HeII reionization extends beyond $z \sim 4$, but
is complete by around then.}
According to this work, HeII$\rightarrow$HeIII reionization
completes within a relatively short redshift range (see the left dashed curve
in the top panel of Fig. \ref{fig:bubbles}).  We approximate
their probability distribution of (late) HeII
reionization redshifts as a Gaussian centered at $4.1$ with a
full-width-at-half-maximum of $0.5$ (dashed curve in the bottom panel
of Fig. \ref{fig:bubbles}).  The quasar spectrum is as described in \S
\ref{Jnu}.

\begin{figure}[tb]
\centerline{\epsfxsize=9cm\epsffile{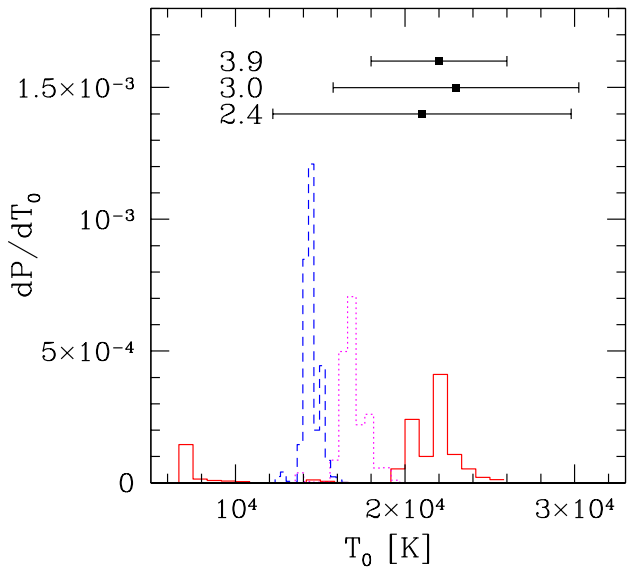}}
\caption{Probability distribution of temperature $T_0$ for fluid elements at
mean density, for three different redshifts $z = 2.4$ (blue dashed histogram), $3.0$
(magenta dotted histogram), and $3.9$ (red solid histogram). 
The observed temperatures with their 2 $\sigma$ errorbars are shown at the top.
}
\label{probT}
\end{figure}

We generate reionization histories for an ensemble of fluid elements
(at mean density) in a stochastic way: determining the redshifts of HI
(and HeI) reionization, and HeII reionization by drawing from the
probability distributions described in the bottom panel of
Fig. \ref{fig:bubbles}. The resulting temperatures at $z = 2.4$,
$3.0$, and $3.9$ have a scatter because of the stochastic history, and
their probability distributions are shown in Fig. \ref{probT}.

The results shown in this figure are intriguing. At $z = 3.9$, the
distribution (red solid histogram) peaks around $T_0 = 2.2 \times 10^4$ K,
but it has a significant bump below $10^4$ K as well.  This bimodal
distribution is a result of the stochastic reionization history: at $z
= 3.9$, there is a minority of fluid elements that have not undergone
HeII reionization, and so they are significantly colder, by more than
a factor of $2$.  There have been attempts in the past to look for
temperature fluctuations (beyond what one expects from variations with
density) in the Ly$\alpha$ forest (Zaldarriaga 2002, Theuns et
al. 2002b). So far there has been no detection.  
Our results indicate that the stochasticity
of reionization can detectably increase the width of the temperature
distribution.  It will be very interesting to confirm our results by a
more detailed analysis that folds in a realistic density distribution,
and to apply the observational search techniques to larger datasets
and to higher redshifts ($z \sim 3$ so far).

The formal mean and rms scatter of $T_0$ at $z = 3.9$ is $1.8 \times
10^4 \pm 7 \times 10^3$ K. It can be seen from Fig. \ref{probT} that
the scatter gets progressively smaller as one goes to lower redshifts:
at $z = 3.0$, $T_0 = 1.7 \times 10^4 \pm 8 \times 10^2$ K; at $z =
2.4$, $T_0 = 1.5 \times 10^4 \pm 5 \times 10^2$ K. The scatter at
these lower redshifts are no larger than the expected scatter from
shock-heating as well as dynamics (e.g. Croft et al. 1997, Hui \&
Gnedin 1997, Dave \& Tripp 2001).

\section{Discussion}
\label{discuss}

We find that the temperature of the IGM, as inferred from Ly$\alpha$
forest spectra at redshifts $z\approx 2-4$, leads to several
interesting conclusions.  Especially interesting are the conclusions
we obtain when the Ly$\alpha$ forest temperature is considered
together with the {\it WMAP} results. Our main conclusions are
summarized as follows:

\begin{itemize}

\item {\it A vanilla reionization model, where $X_{\rm HI}, X_{\rm
HeI}$, and/or $X_{\rm HeII}$ undergo order unity changes at $z > 10$,
but suffer no such changes at $z < 10$, is ruled out by temperature
measurements of ZHT01, especially at $z = 3.9$.}  This relies on
assumptions about the ionizing spectrum, which are discussed in \S
\ref{Jnu}.  For a power-law spectrum, rigorous requirements can be put
on the hardness of the ionizing background to evade the the above
argument: $\alpha < 0.12$ if helium is doubly ionized, or $\alpha <
-2.2$ if helium is singly ionized, for $J \propto \nu^{-\alpha}$ (with
a cutoff at $\nu_{\rm HeII}$ if helium is only singly ionized). The
idea of an asymptotic evolution for the thermal state of the IGM is
quite useful in formalizing the above argument. The asymptote (at $2
\le z \le 4$) is described quite accurately ($\sim 5 \%$) by
eq. (\ref{analytic}). It is reached as long as any order unity changes
in $X_{\rm HI}, X_{\rm HeI}$, and/or $X_{\rm HeII}$ occur prior to $z
= 10$.

\item {\it Conversely, the requirement by WMAP that the universe
reionizes early, at $z > 10$, implies the reionization history is
complex. To fulfill the temperature constraints, there must be
additional periods of order unity changes in one or more of the
fractions $X_{\rm HI}, X_{\rm HeI}$, $X_{\rm HeII}$ at redshift below
10.} This is an argument separate from the argument based on the
evolution of mean transmission (Haiman \& Holder 2003; based on the
Gunn-Peterson troughs observed by Becker et al. 2001 and Fan et
al. 2003 at $z \sim 6$, and comparison against an extrapolation from
lower redshifts).

\item {\it Exactly what kind of changes at $z < 10$ are necessary to
match the temperature constraint at $z = 3.9$ depends critically on
the ionizing spectrum.} If the spectrum is harder than $\nu^{-1.5}$
above the HI and HeI ionization thresholds (i.e. harder than the
'quasar' spectrum discussed in \S \ref{Jnu}), then order unity changes
in $X_{\rm HI}$ and $X_{\rm HeI}$ in the period $6 \lsim z \lsim 10$
(the lower limit of $6$ being set by the SDSS mean transmission
measurements; Fan et al. 2002), even without accompanying changes in
$X_{\rm HeII}$ (i.e. no HeII reionization), are sufficient to heat up
the IGM to within the constraint at $z = 3.9$.  Conversely, if the
spectrum is not hard enough (for instance, a population II stellar
spectrum described in \S \ref{model}), some amount of HeII
reionization-heating is {\it necessary} prior to $z = 3.9$.

\item {\it The IGM temperature can have a broad distribution
(beyond what one expects based on variation with density)
close to reionization events, but the scatter diminishes with time.}
The reionization history of the universe is almost
certainly stochastic, in the sense that different fluid elements
reionize at different times, depending on when a given element becomes
engulfed in expanding HII (or HeIII) regions around ionizing sources.
Based on simple semi-analytic models, we predict the probability
distribution of ionization redshifts.  The resulting temperature
scatter is particularly large close to reionization events, but
diminishes quickly with time. This is illustrated in Fig. \ref{probT}.
It would be very interesting to search for the broad initial
temperature scatter that is predicted by physically motivated
reionization models.

\item {\it The known quasar population at $z<5$ may heat the IGM to
sufficiently high temperatures by reionizing HeII.}  In \S
\ref{model}, we describe a physically motivated reionization model
with a stochastic history. Hydrogen (and HeI) reionization is
accomplished early on by metal-free stars ($17 \lsim z \lsim 30$), and
later on by population II stars ($z \lsim 17$), and completes by $z
\sim 6.5$.  This fulfills the dual requirements of a large CMB optical
depth, and the Gunn-Peterson trough seen by SDSS. Because the
population II stars have a rather soft spectrum, photo-heating of HI
and HeI alone is not enough to bring the temperature up to the
observed level, and this model {\it fails} to predict sufficiently
high IGM temperatures -- despite the percolation occurring at the
relatively low redshift of $z\sim 7$.  We therefore make use of the
predictions of Sokasian et al. (2002) for a brief period of HeII
reionization around $z \sim 4$, based on the {\it known} population of
quasars. We find that such a model can satisfy the temperature
constraints.  It is important to note, however, that this is not the
only way to achieve the observed temperatures. For instance, a
population of dim quasars can turn on at $z > 6$ (Haiman \& Loeb
1997), which has a sufficiently hard spectrum to either heat the IGM
via photo-heating of HI and HeI alone ( but their spectrum has to be
harder than $\nu^{-1.5}$ if HeII is not reionized), or boost the
temperature via HeII reionization as well. If HeII reionization takes
place via these mini-quasars, the turn-on of the known population of
quasars at $z \lsim 4$ then has a much weaker influence on the thermal
state at low redshifts.

\item {\it HeII reionization makes little difference to the electron
scattering optical depth, particularly if it occurs late.}
For instance, in the model shown in
Fig. \ref{fig:bubbles}, extra electrons from HeII reionization changes
$\tau_{e}$ at a $2 \%$ level fractionally.

\end{itemize}

An important caveat in our discussion so far is the possible existence
of additional heating mechanisms.
Two such possibilities are galactic
outflows and Compton heating by a hard X-ray background.  Adelberger
et al. (2002) recently observed signatures of galactic winds into the
IGM.  Such outflows can in principle heat up the IGM. However,
observations by Rauch et al. (2001) of close pairs of lines of sight
in lens systems suggest that the IGM is not turbulent on small-scales,
arguing against significant stir-up of the IGM by winds.  Moreover,
galactic outflows can heat up the IGM to a variety of temperatures --
the fact that the observed temperatures are in the range of
expectations for a photo-ionized, and photo-heated, gas suggests
photo-heating is the simplest explanation.

Another important question is whether Compton heating by a hard X-ray
background (XRB) could be more important than photoelectric heating.
This question was considered by Madau \& Efstathiou (1999), who
assumed that the redshift evolution of the sources of the hard XRB
parallels the flat distribution that had been determined for the soft
X--ray AGN luminosity function beyond $z\sim 2$. Under this
assumption, they found the hard XRB to be an important source of
heating relative to the UV background at redshifts $z>2$, raising the
IGM temperature to $1.5\times10^4$K at $z\sim 4$.  Two new
developments make it unlikely for the hard XRB to be an important
source of heating.  First, as pointed out by Abel \& Haehnelt (1999),
the photoelectric heating rate can be significantly increased in
optically thick gas (when hydrogen and helium first gets ionized); we
here adopt these increased rates.  Second, the sources of the hard XRB
have been resolved by the {\it Chandra} satellite, and, rather than
paralleling the soft X--ray luminosity function, the hard X--ray
sources exhibit a steep decline towards high redshift beyond $z\sim 2$
(Cowie et al. 2003).

Our investigation raises a number of interesting issues.  Is HeII
reionization prior to redshift $4$ necessary to heat up the IGM to the
right level?  This depends critically on what kind of sources and
spectra are available.  The prediction for a large scatter in
temperature (factor of about $2$) close to the epoch of HeII
reionization is something one could look for, if indeed HeII
reionization occurs late (Zaldarriaga 2002, Theuns et al. 2002b).  In
this paper, we have focused entirely on the thermal state of fluid
elements at the mean density.  The formalism of Hui \& Gnedin (1997)
can be used to compute the same for elements at $\delta\rho/\rho \lsim
10$.  It is interesting to explore how the variation of temperature
with density ($T \propto \rho^{\gamma -1 }$, an effective equation of
state) might place additional constraints on the reionization history.
At present, measurements of $\gamma$ are quite noisy (e.g. ZHT01).
New approaches to constrain it better will therefore be
very useful (Dijkstra, Lidz \& Hui 2003).
Finally, there is the issue of exotic ionization mechanisms (e.g.
Berezhiani \& Khlopov 1990) which is particularly interesting
in the face of early reionization.

We thank Renyue Cen, Andrea Ferrara, Wayne Hu, Avi Loeb, Jerry Ostriker, 
Joop Schaye, and
especially Adam Lidz, for useful discussions.
LH is supported in part by an Outstanding Junior Investigator Award
from the DOE, an AST-0098437 grant from the NSF, and by the DOE at
Fermilab, and NASA grant NAG5-10842.




\end{document}